\documentclass[fleqn,usenatbib,useAMS]{mnras}

\usepackage{graphicx}	% Including figure files
\usepackage{amsmath}	% Advanced maths commands
\usepackage{amssymb}	% Extra maths symbols
\usepackage{multicol}        % Multi-column entries in tables
\usepackage{bm}		% Bold maths symbols, including upright Greek
\usepackage{pdflscape}	% Landscape pages
\usepackage{framed,multirow}
\usepackage{booktabs}
\usepackage{latexsym}
\usepackage{url}
\usepackage{xcolor}

\usepackage{hyperref}
\DeclareMathAlphabet{\mathcal}{OMS}{cmsy}{m}{n}

\usepackage[T1]{fontenc}
\usepackage{ae,aecompl}

\usepackage{mathptmx}
\title[Anonymized labeled TOF-MRA images using GANs]{Anonymization of labeled TOF-MRA images for brain vessel segmentation using generative adversarial networks}

\author[T. Kossen et al.]{Tabea Kossen$^{1,2}$,
Pooja Subramaniam$^{1,3}$,
Vince I Madai$^{1,4}$,
Anja Hennemuth$^{2,5,6}$,
Kristian Hildebrand$^{7}$,
\newauthor Adam Hilbert$^{1,2}$,
Jan Sobesky$^{8,9}$,
Michelle Livne$^{1}$,
Ivana Galinovic$^{9}$,
Ahmed A Khalil$^{9,10,11,12}$,
\newauthor Jochen B Fiebach$^{9}$ and
Dietmar Frey$^{1}$ 
\\
$^{1}$CLAIM - Charit\'{e} Lab for AI in Medicine, Charit\'{e} Universit\"{a}tsmedizin Berlin, Germany \\
$^{2}$Department of Computer Engineering and Microelectronics, Computer Vision \& Remote Sensing, Technical University Berlin, Berlin, Germany \\
$^{3}$Department of Electrical Engineering and Computer Science, Technical University of Berlin, Berlin, Germany \\
$^{4}$School of Computing and Digital Technology, Faculty of Computing, Engineering and the Built Environment, Birmingham City University, Birmingham, UK \\
$^{5}$Institute for Imaging Science and Computational Modelling in Cardiovascular Medicine, Charit\'e Universit\"atsmedizin Berlin, Berlin, Germany \\
$^{6}$Fraunhofer MEVIS, Bremen, Germany \\
$^{7}$Department VI Computer Science and Media, Beuth University of Applied Sciences, Berlin, Germany \\
$^{8}$Johanna-Etienne-Hospital, Neuss, Germany \\
$^{9}$Centre for Stroke Research Berlin, Charit\'e  Universit\"atsmedizin Berlin, Berlin, Germany \\
$^{10}$Department of Neurology, Max Planck Institute for Human Cognitive and Brain Sciences, Leipzig, Germany \\
$^{11}$Mind, Brain, Body Institute, Berlin School of Mind and Brain, Humboldt University Berlin, Berlin, Germany \\
$^{12}$Berlin Institute of Health, Berlin, Germany}

\pubyear{2020}

\begin{document}
\label{firstpage}
\pagerange{\pageref{firstpage}--\pageref{lastpage}}
\maketitle

\begin{abstract}
Anonymization and data sharing are crucial for privacy protection and acquisition of large datasets for robust medical image analysis. This represents a major challenge, especially for brain imaging research. Here, the unique structure of brain images allows for potential re-identification and thus requires anonymization beyond conventional methods. Generative adversarial networks (GANs) have the potential to provide anonymous images while maintaining their predictive properties.

Analyzing brain vessel segmentation, we trained 3 GAN architectures on time-of-flight (TOF) magnetic resonance angiography (MRA) patches of patients with cerebrovascular disease for image-label pair generation: 1) Deep convolutional GAN, 2) Wasserstein-GAN with gradient penalty (WGAN-GP) and 3) WGAN-GP with spectral normalization (WGAN-GP-SN). First, the synthesized image-labels from each GAN architecture were used to train a U-net for vessel segmentation. The U-nets were then tested on real patient data. In total, 66 patients were used for this analysis. In a second step, we simulated the application of our synthetic patches in a transfer learning approach using a second, independent dataset. Here, for an increasing number of up to 15 patients we evaluated vessel segmentation model performance on real data with and without pre-training on generated patches. Finally, performance for all models was assessed by the Dice Similarity Coefficient (DSC) and the 95th percentile of the Hausdorff Distance (95HD).

Comparing the 3 GAN architectures, the U-net model trained on synthetic data generated by the WGAN-GP-SN showed the highest performance to predict brain vessels (DSC/95HD 0.82/28.97) benchmarked by the U-net trained on real data (0.89/26.61). The transfer learning approach showed superior performance for the same GAN architecture compared to no pre-training, especially for one labeled patient only (DSC/95HD 0.91/25.68 compared to DSC/95HD 0.85/27.36).

In a brain imaging segmentation paradigm, synthesized image-label pairs preserved generalizable information and showed good performance for vessel segmentation. Furthermore, we showed that synthetic patches can be used in a transfer learning approach with an independent dataset. These results pave the way to overcome the crucial challenges of scarce data and anonymization in the medical imaging field. To facilitate further research, our synthetic image-label pairs are being made available upon request.

\end{abstract}


%%%%%%%%%%%%%%%%%%%%%%%%%%%%%%%%%%%%%%%%%%%%%%%%%%

%%%%%%%%%%%%%%%%% BODY OF PAPER %%%%%%%%%%%%%%%%%%

% The MNRAS class isn't designed to include a table of contents, but for this document one is useful.
% I therefore have to do some kludging to make it work without masses of blank space.
\begingroup
\let\clearpage\relax
%\tableofcontents
\endgroup
\newpage

\section{Introduction}

Modern deep learning methods have revolutionized the field of natural image analysis (\cite{krizhevsky_imagenet_2017, simonyan_very_2014}). These methods are translated to medical image analysis with growing success (\cite{litjens_survey_2017, navab_u-net:_2015, livne_u-net_2019}). However, in contrast to natural images, the number of data sets in medical image analysis are usually orders of magnitude smaller since their availability is limited owing to data privacy regulation. This poses a continuous challenge for deep learning research in the medical imaging field. To meet this challenge, anonymization of medical images is an essential method to ensure both data privacy and data availability for research. However, current anonymization methods in neuroimaging such as face blurring or face removal still allow re-identification and thus cannot be applied (\cite{abramian_refacing_nodate, ravindra_-anonymization_2019, wachinger_brainprint_2015}). These results call for new techniques to anonymize medical neuroimaging data to both protect patient privacy and to facilitate research progress. 

Generative adversarial networks (GANs) have the potential to fulfill this need. GANs have already been applied successfully for medical imaging data synthesis (\cite{neff_generative_nodate, yi_generative_2019, sorin_creating_2020}). Also, first pilot studies have already made use of GANs for anonymization purposes (\cite{shin_medical_2018, bebis_deepprivacy_2019}). However, applications for neuroimages are scarce and synthesizing images often requires additional patient information such as a segmentation label (\cite{shin_medical_2018}). This means that patient information is still fed into the model and the generated images are then not properly anonymized. Thus, there is a need to investigate the ability of GANs to create state-of-the-art anonymous synthetic neuroimaging data maintaining the predictive properties of the original data. Importantly, such an approach would have the most beneficial impact if the corresponding labels would be created in the same process since many supervised deep learning applications require time-consuming manual labeling of the dataset by experienced physicians. 

In this work, we utilize arterial brain vessel segmentation to test the ability of GANs to create synthetic neuroimaging data and corresponding labels. Moreover, we investigate the generalizability of the synthesized data on a second, independent dataset. With respect to the generative architectures, we train 3 different GAN architectures on time-of-flight (TOF) magnetic resonance angiography (MRA) image patches of patients with cerebrovascular disease: 1) Deep Convolutional GAN (DCGAN), 2) Wasserstein GAN with gradient penalty (WGAN-GP) and 3) WGAN-GP using spectral normalization (WGAN-GP-SN). With each GAN type, we synthesized both the image and the corresponding label. We validate the generated synthetic patches using two different approaches. In the first approach, we evaluate the quality of the generated patches a) using the Fr\'{e}chet inception distance (FID) and b) by training a vessel segmentation U-net on the synthetic patches. The U-net's performance is then assessed on real test data. In total, 66 patients were utilized for this analysis. In the second approach, we use the synthetic patches to pre-train a vessel segmentation model and apply the network weights in a transfer learning setting to pre-initialize the training of a U-net model using up to 15 patients from a second, independent TOF-MRA dataset. The performance of this model is then compared to a U-net model without any pre-training. Finally, to facilitate and accelerate future research on arterial vessel segmentation and to corroborate the usefulness of the effective anonymization procedure, we make the synthesized image-label pairs generated in our study available upon request. 

Taken together, the contributions of the paper are: We present effectively anonymized and labeled TOF-MRA patches for brain vessel segmentation for the first time to our knowledge. Furthermore, we compare three different state-of-the-art GAN architectures and evaluate our synthesized labeled data on an independent, second dataset in a novel evaluation pipeline. We show that pre-training a vessel segmentation network using our synthetic data yields superior performance compared to no pre-training and can reduce the amount of additional training data. Finally, we make our synthesized data available upon request to facilitate further research.

\section{Related Work}

GANs have already been shown to be successful in many applications of data augmentation in medical imaging (\cite{frid-adar_gan-based_2018}, \cite{sandfort_data_2019}) as well as in neuroimaging (\cite{bowles_gan_2018}, \cite{foroozandeh_synthesizing_2020}). Here, real medical images together with synthesized images were used to improve models that were trained on real data only. Whereas we provide results on data augmentation, this study focussed on the models trained on purely synthetic data and its generalizability to a new dataset. 

Generating medical images with labels is not a new idea. \cite{neff_generative_nodate} showed that lung x-rays with corresponding segmentation labels can be generated using a GAN architecture. \cite{guibas_synthetic_2018} demonstrated the synthesization of labeled retina images using two GANs. While these studies focused on 2D medical images, we use a 3D dataset and evaluate the performance on an independent dataset. 
In the neuroimaging domain \cite{foroozandeh_synthesizing_2020} recently showed that synthesized and labeled MR images can improve tumor segmentation performance. However, the focus here was on augmentation and models trained on synthesized data alone yielded comparably low performance. Also, in contrast to the present study, only one dataset was used for training and evaluation.

\section{Methods}

\subsection{Network Architecture}
\begin{figure*}[!t]
	\centering
	\includegraphics[width=\textwidth]{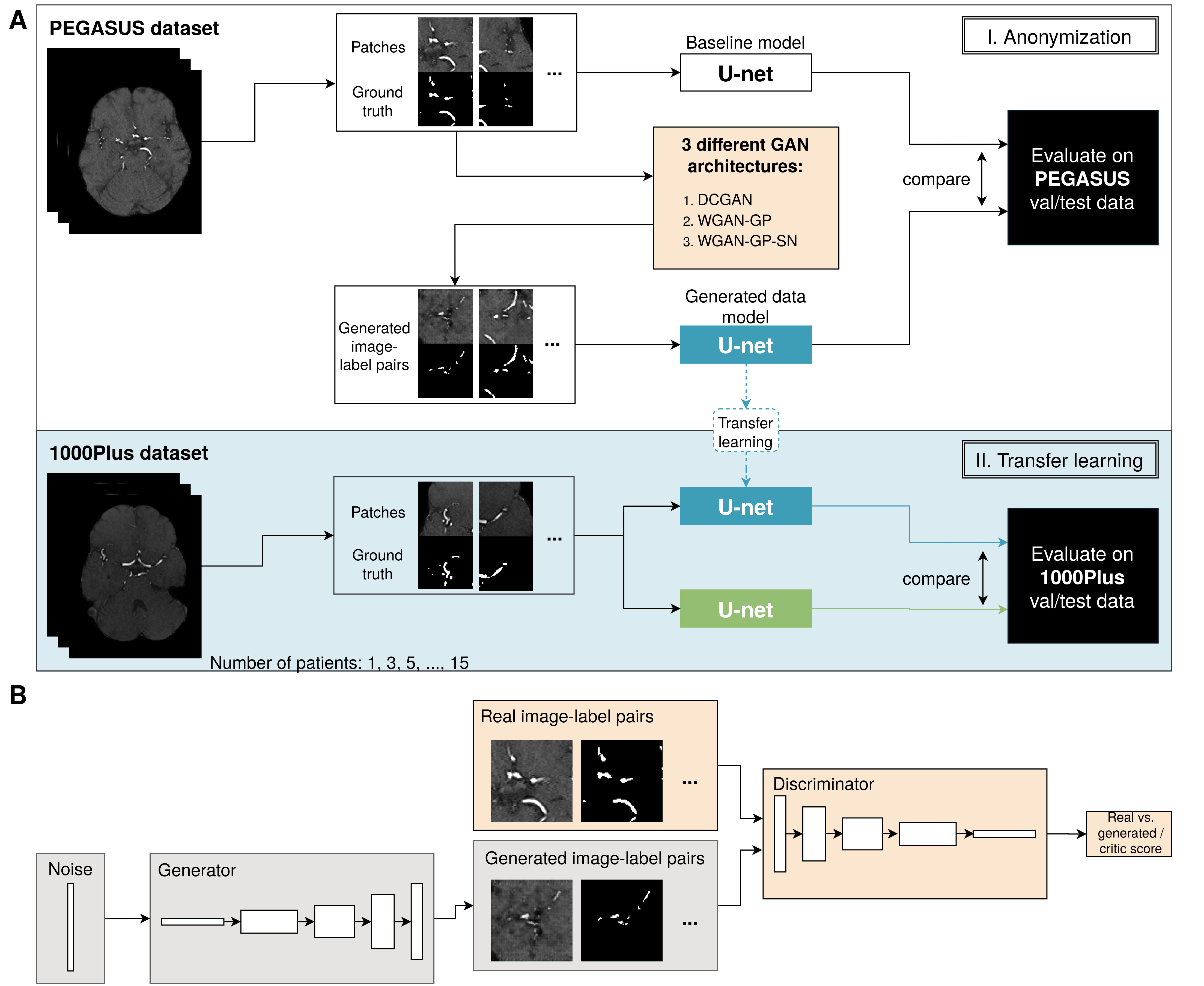}
	\caption{ Workflow of this study (A) and basic architecture of the generative adversarial networks that were trained (B).} \label{fig1}
\end{figure*}

The architecture of the proposed DCGAN was adapted from \cite{radford_unsupervised_2016} and \cite{neff_generative_nodate}.  The WGAN-GP is an extension of the original Wasserstein GAN  (\cite{arjovsky_wasserstein_2017}) using gradient penalty for regularization (\cite{gulrajani_improved_nodate}). For the third architecture WGAN-GP-SN spectral normalization was used in the convolutional layers of the WGAN-GP (\cite{miyato_spectral_2018}). Our code is available at \href{https://github.com/prediction2020/GANs-for-anonymized-labeled-TOF-MRA-patches}{https://github.com/prediction2020/GANs-for-anonymized-labeled-TOF-MRA-patches}. The proposed methods and the structure of the GAN is shown in Fig.~\ref{fig1}. 

The generator $G$ of all architectures took a noise vector of length 100 sampled from a gaussian distribution as input. The noise vector was then fed through 6 upsampling convolutional layers using a kernel size of 5 and stride of 2. After each convolution layer, a batch normalization layer and a ReLU activation layer were added, except for the last convolution layer. The activation function used after the last convolution layer is the hyperbolic tangent function. The network then outputs two 96 x 96 images that correspond to one image-label pair $x_{\text{gen}} \sim p_{gen}$. The objective function for the generators of all architectures were built upon:

\begin{equation}
\mathcal{L}_{G} = \text{max}_{G}  
\mathbb{E}_{x_{\text{gen}} \sim p_{\text{gen}}} [log(D(x_{\text{gen}}))]	
\end{equation}

The discriminator $D$ for all architectures took two 96 x 96 images as input which correspond to either a real image-label pair or generated image-label pair. The pairs were again fed through 6 convolutional layers with a kernel size of 5 and stride of 2. After each convolution layer, a batch normalization layer and a leaky ReLU (with a slope of 0.2) were added, except for the last convolution layer. The activation function used after the last convolution layer in the DCGAN was a sigmoid function. The objective function of the discriminator for the DCGAN was:

\begin{equation}
\mathcal{L}_{D} = \text{max}_{D} \mathbb{E}_{x_{\text{real}} \sim p_{\text{real}}} [logD(x_{\text{real}})] +  
\mathbb{E}_{x_{\text{gen}} \sim p_{\text{gen}}} [log(1-D(x_{\text{gen}}))]
\end{equation}

where $x_{\text{real}} \sim p_{\text{real}}$ denoted the real image-label pair.

For the WGAN-GP and WGAN-GP-SN, a gradient penalty term for regularization was added to the discriminator's loss:

\begin{equation}
\text{loss}_{D} = D(x_{\text{gen}}) - D(x_{\text{real}}) + \lambda (\lVert \nabla D(\epsilon x_{\text{real}} + (1-\epsilon) x_{\text{gen}}) \rVert - 1 )^{2},
\end{equation}

where $\epsilon \sim \textit{U}[0,1]$ and $\lambda=10$. Since the discriminator acted as a critic, the sigmoid activation function in the last convolutional layer was omitted. The batch normalization was replaced by instance normalization to normalize across features and channels in the WGAN-GP. In the WGAN-GP-SN architecture, spectral normalization was used instead of instance normalization.

For training the DCGAN, the Adam optimizer (\cite{kingma_adam_2017}) with a learning rate of 0.0003 with $\beta_1=0.5$ was used for both the generator and the discriminator. The batch size was 512 and the model was trained for 178 epochs. To improve stability of the training, label smoothing (ranges 0.7-1.2/0-0.3) and feature matching between the last convolutional layer using L1 norm were applied (\cite{salimans_improved_2016}). 

For WGAN-GP and WGAN-GP-SN, the Adam optimizer was utilized with a learning rate of 0.0001 with $\beta_1=0$ and $\beta_2=0.9$ for both generator and discriminator. The batch size was 300 and both models were trained for 180 epochs. In each epoch the discriminator was updated five times and the generator once.
All models were implemented in PyTorch and trained on two GeForce GTX 1080Ti.

\subsection{Patients}
A total of 121 patient MRA data from two studies were used: PEGASUS (N=66) and 1000Plus (N=55). All patients were diagnosed with a cerebrovascular disease. Details on both studies can be found in previous papers, for the PEGASUS study see (\cite{mutke_clinical_2014}), for the 1000Plus study see  (\cite{hotter_prospective_2009}). All the patients gave their informed written consent. The studies have been conducted in accordance with the authorized ethical review committee of Charit\'e - Universit\"{a}tsmedizin Berlin.  

Scans were performed on a clinical 3T whole-body system (Magnetom Trio, Siemens Healthcare, Erlangen, Germany; using a 12-channel receive radiofrequency coil (Siemens Healthcare) tailored for head imaging.

Parameters PEGASUS: voxel size = (0.5x0.5x0.7) mm$^3$; matrix size: 312x384x127; TR/TE = 22ms/3.86ms; acquisition time: 3:50 min, flip angle = 18 degrees. 

Parameters 1000Plus: voxel size = (0.5x0.7x0.7) mm$^3$; matrix size: 384x268x127; TR/TE = 22ms/3.86ms; acquisition time: 3:50 min, flip angle = 18 degrees. 

For both datasets, skull-stripping was applied. The segmentation labels were produced semi-manually using a standardized pipeline along with 4 raters correcting the labels as described in \cite{livne_u-net_2019}. 

\subsection{Data Splitting and Patch Extraction}
For the anonymization, 41 out of the 66 PEGASUS patients were used as a training set, 11 were used for validation and 14 for testing. For the transfer learning approach, one to 15 patients in increments of two of the 1000Plus data were utilized for training. The 1000Plus validation set consisted of 10 and the test set of 40 patients.

Due to memory considerations, 2D patches of size 96x96 were extracted from each patient instead of using the whole volume. The data contained 1\% vessels and 99\% background. To compensate for this imbalance, 500 patches per patient with a brain vessel in the center were extracted. Then, 500 random patches per patient were added. The input patches were normalized to a range between -1 and 1 for the GAN used for anonymization. For the U-net segmentation model, the input was normalized patch-wise to zero-mean and unit-variance.

\subsection{Performance Evaluation}
The generated images were first visually inspected and then quantitatively compared to the real data using the Fr\'echet inception distance (FID) (\cite{heusel_gans_2018}). The FID measures the similarity of the real and generated images by feeding both into an Inception-v3 network. The difference between the activations in the pool3 layer inside the Inception-v3 network is then calculated as follows:

\begin{equation}
\text{FID} = \parallel \mu_{real} - \mu_{gen} \parallel^{2} + \text{Tr}(\sigma_{real} + \sigma_{gen} - 2(\sigma_{real}\sigma_{gen})^{1/2}),
\end{equation}

where $x_{real} \sim \mathcal{N}(\mu_{real},\sigma_{real})$ and $x_{gen} \sim \mathcal{N}(\mu_{gen},\sigma_{gen})$ are the distributions of the features in the pool3 layer of the real and generated data respectively. 

The FID was calculated for 41,000 generated patches of all three architectures with the respective 41,000 real patches. The lower the FID, the higher the similarity of the generated data to the original data.

As a second evaluation, the state-of-the-art "half U-net" used in \cite{livne_u-net_2019} was trained with generated data as well as both real and generated data. The parameters learning rate and dropout rate were tuned with respect to the validation set. Additionally, classical augmentation was used as described in \cite{livne_u-net_2019} if this led to an improved performance on the validation set. Each segmentation network was trained for 15 epochs. Then, the performance was evaluated on the binary segmentation maps of the test set by the DSC and the 95th percentile of the Hausdorff distance (95HD):

\begin{equation}
\text{DSC} = \frac{2\text{TP}}{2\text{TP} + \text{FP} + \text{FN}},
\end{equation}

where TP are the true positives, FP the false positives and FN the false negatives. The Hausdorff distance is defined as:

\begin{equation}
\text{HD} = \text{max}(\text{max}_{i\in[0,N-1]} d(i,P,G), \text{max}_{i\in[0,M-1]} d(i,G,P)),
\end{equation}

where $N$ and $M$ denote the number of voxels on the vessel tree of the ground truth $G$ and the prediction $P$ respectively. $d(i,P,G)$ is defined as the distance from vessel voxel $i$ in $G$ to the closest vessel voxel in $P$. The 95HD was then the 95th percentile Hausdorff distance for each voxel, averaged over each voxel and each patient. It was measured in millimeters.

In the second part of the analysis, the performance of the U-net trained on generated patches was evaluated on the 1000Plus dataset. For an increasing number of training patients (1, 3,\ldots , 15) the U-net was trained from scratch and using the weights from the model trained on the generated image-label pairs (transfer learning). The performance of using real data only and transfer learning was then compared by assessing the DSC and 95HD on the validation (10 patients) and test set (40 patients).

\section{Results}

\begin{figure*}
	\centering
	\includegraphics[width=0.9\textwidth]{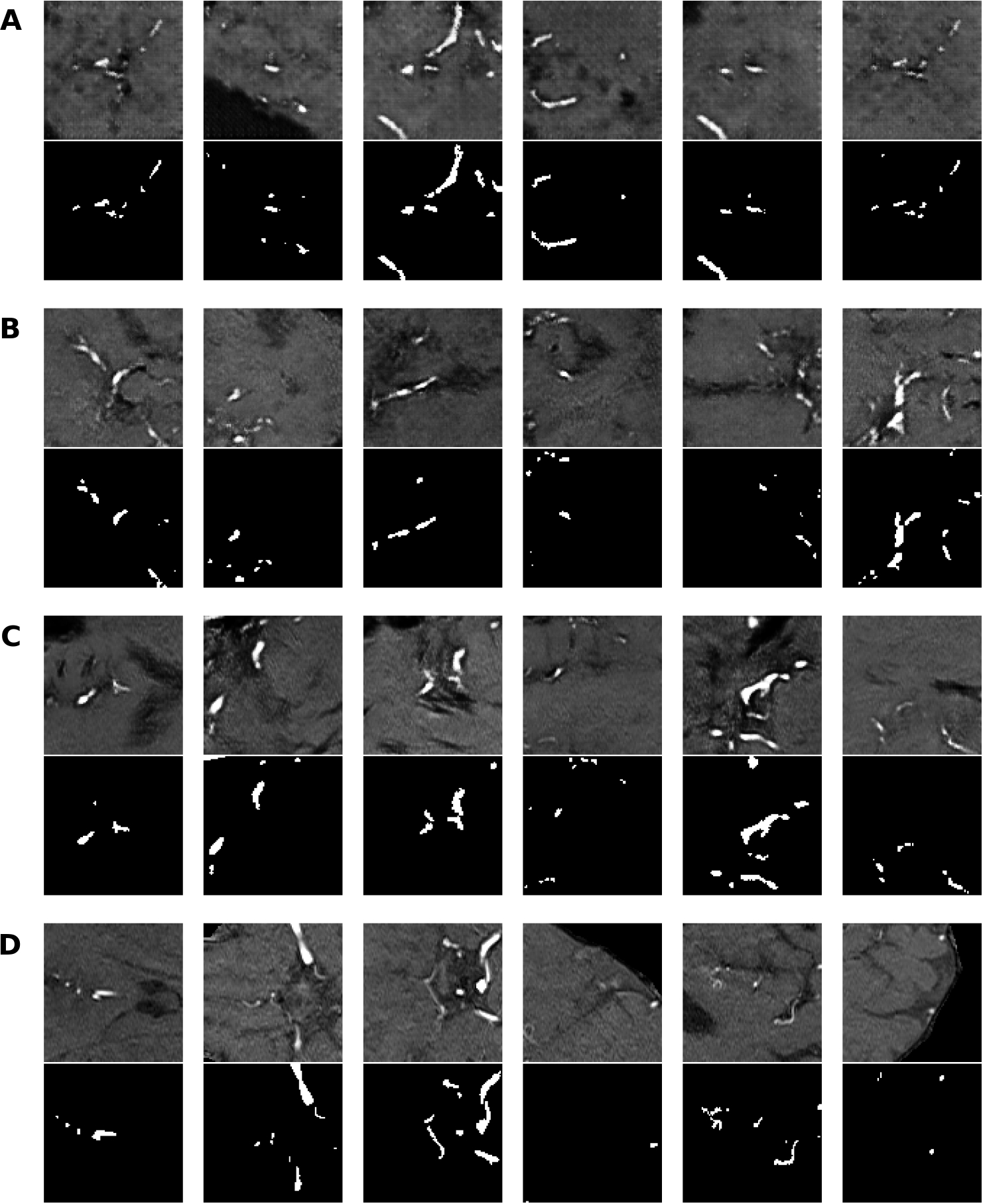}
	\caption{Real and synthesized image patches with corresponding labels. (A) to (C) show image-label pairs generated by DCGAN (A), WGAN-GP (B) and WGAN-GP-SN (C) respectively. (D) show real patches and corresponding labels. The synthesized patches resemble real vessel patches and the labels fit well to the patches, especially those generated by WGAN-GP-SN (C).} \label{fig2}
\end{figure*}

Overall, generated synthetic patches showed high similarities to the training set patches, in particular those that were synthesized by the WGAN-GP-SN. The patches generated by the DCGAN showed a lower resolution with slight checkerboard artifacts compared to the original patches. The generated corresponding labels fit well to the patches for all models. A subset of the synthesized image-label pairs for all GAN architectures as well as original image-label pairs are shown in Fig.~\ref{fig2}A to D. In the quantitative assessment, the data generated by the WGAN-GP-SN architecture showed the highest similarity to the real data with a FID of 38.05 compared to 105.96 for the worst performing DCGAN. All FID values for real and synthesized data can be found in Table~\ref{tab1}.

\begin{table}
	\caption{Fr\'{e}chet inception distance (FID) as a quantitative measurement of the generated image's similarity compared to the real images for each of the three GAN architectures. WGAN-GP-SN showed the highest similarity to the real data in terms of FID.}\label{tab1}
	\centering
	\begin{tabular}{lc}
		\toprule
		GAN architecture & FID \\
		\midrule
		DCGAN & 105.96 \\
		WGAN-GP & 52.53 \\
		WGAN-GP-SN & \textbf{38.05} \\
		\bottomrule
	\end{tabular}
\end{table}

In the first validation approach, The U-net trained on data generated by the WGAN-GP-SN showed the highest performance of all GAN models with a segmentation performance of 0.82 DSC/28.97 95HD. The U-net trained on real PEGASUS data showed a performance of 0.89 DSC/26.57 95HD. The same model showed a similarly high performance in the external validation on the 1000Plus data with 0.88 DSC/25.68 95HD. Quantitative results for all models trained on generated and/or real data can be found in Table~\ref{tab2}.

\begin{table*}
	\caption{Summary of the Dice similarity coefficient (DSC) and the 95th-percentile Hausdorff distance (95HD) of the U-net on validation and test set. Each metric is averaged over patients. The artificial patches were generated by Generative Adversarial Networks (GANs) trained on the PEGASUS dataset. For data augmentation, both real and generated patches have been used for training. For anonymization the U-net was trained on generated patches only. Models trained on anonymized, synthetic data only show performances close to the model trained on real data.}\label{tab2}
	\centering
	\begin{tabular}{lcccc}
		\toprule
		\multirow{2}{*}{}                                             & \multicolumn{2}{c}{mean DSC} & \multicolumn{2}{c}{mean 95HD {[}mm{]}} \\
		& val          & test          & val               & test               \\
		\midrule
		U-net on real PEGASUS data (Livne et al.)                          & 0.88           & 0.89          & 29.50               & 26.57                \\
		\midrule
		\multicolumn{1}{@{}l}{\textbf{Data augmentation (real data (PEGASUS) and generated data)}} \
		&              &               &                   &                    \\
		%\multicolumn{1}{@{}l}{\textit{PEGASUS data and generated patches}} \
		%&              &               &                   &                    \\
		DCGAN                                                        
		& 0.89         & \textbf{0.90} & 30.22             & \textbf{25.61}       \\
		WGAN-GP
		& 0.89         & 0.89          & 28.03             & 30.01              \\
		WGAN-GP-SN                                                    
		& 0.89         & 0.89          & 29.91             & 26.51     \\
		\midrule
		\multicolumn{1}{@{}l}{\textbf{Anonymization (trained on generated data only)}} \
		&              &               &                   &                    \\
		\multicolumn{1}{@{}l}{\textit{PEGASUS anonymization models: validated and evaluated on PEGASUS data}} \
		&              &               &                   &                    \\
		DCGAN                                                         
		& 0.82         & 0.79          & 34.58             & 31.25              \\
		WGAN-GP                                                       
		& 0.82         & 0.78          & 34.64             & 33.70              \\
		WGAN-GP-SN                                                    
		& 0.85         & \textbf{0.82} & 30.88             & \textbf{28.97}     \\
		%\midrule
		\multicolumn{1}{@{}l}{\textit{PEGASUS anonymization models evaluated on real 1000Plus}} \
		&              &               &                   &                    \\
		%		DCGAN                                                         
		%		& 0.77         & 0.76          & tbd               & tbd                \\
		%		WGAN-GP                                                       
		%		& tbd          & 0.85          & 34.64             & 33.70              \\
		%		WGAN-GP-SN                                                    
		%		& tbd          & \textbf{0.88} &              &               \\
		DCGAN                                                         
		& -            & 0.76          & -                 & 26.79              \\
		WGAN-GP                                                       
		& -            & 0.85          & -                 & 26.98              \\
		WGAN-GP-SN                                                    
		& -            & \textbf{0.88} & -                 & \textbf{25.68}       \\
		\bottomrule
	\end{tabular}
\end{table*}

\begin{figure*}[]
	\centering
	\includegraphics[width=0.95\textwidth]{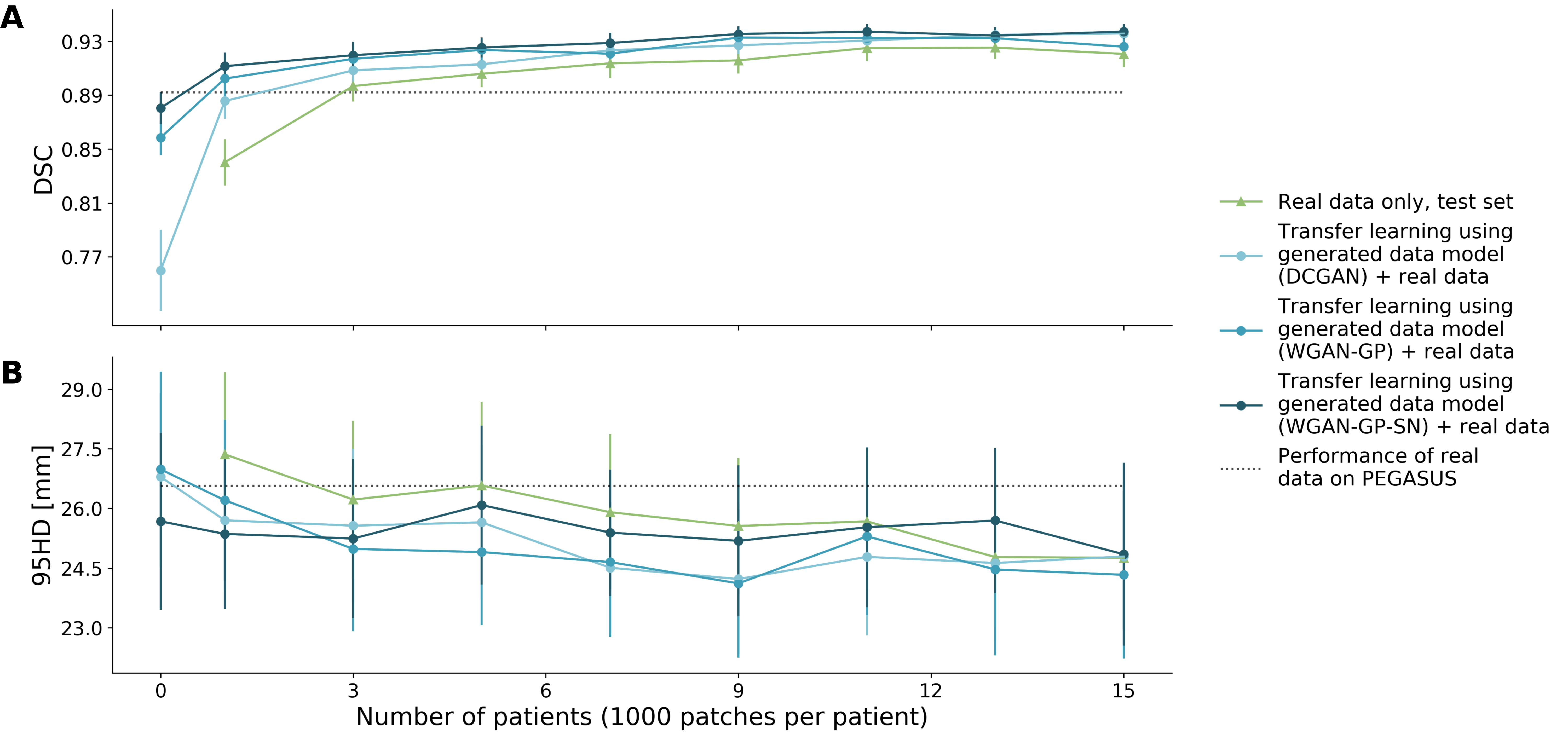}
	\caption{Performance evaluation for segmentation for an increasing number of patients on the 1000Plus dataset when trained from scratch (green) and using transfer learning (blue). The black dotted lines indicate the performance of the Unet on the real PEGASUS dataset. The error bars show the standard deviation over the patients. Especially for up to 5000 data samples the pre-trained WGAN-GP-SN outperform the models without any pre-training.} 
	\label{fig3}
\end{figure*}

\begin{figure*}[]
	\includegraphics[width=\textwidth]{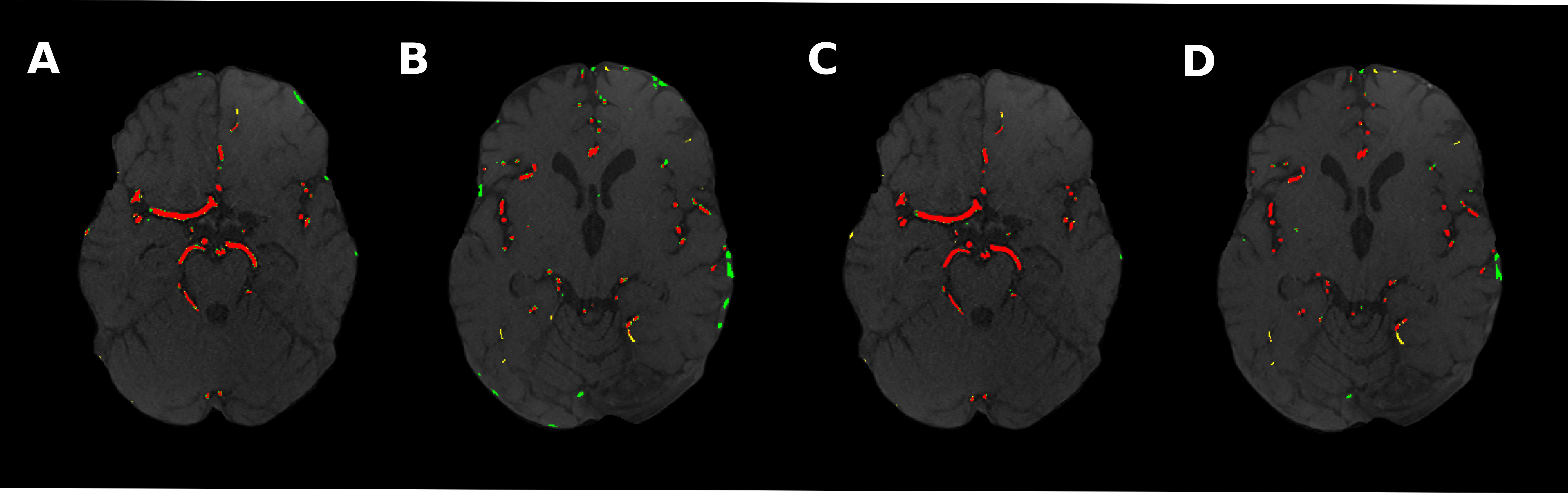}
	\caption{Error maps for one example patient from the 1000Plus study using one patient when training from scratch (A, B) and using transfer learning from WGAN-GP-SN generated patches (C, D). True positives are shown in red, false positives in green and false negatives in yellow. Transfer learning led to less errors, especially on small vessels (B, D).} 
	\label{fig4}
\end{figure*}

In the second validation approach applying transfer learning, the U-net pre-initialized with the weights from training on synthesized patches exhibited a higher performance compared to the model trained from scratch on real data only could be observed. Particularly when training on patches from one patient only (n=1000), transfer learning using patch-label pairs generated by the WGAN-GP-SN led to a higher performance in terms of DSC and 95HD (DSC/95HD 0.91/25.68 compared to 0.85/27.36). This observed performance difference between pre-initialized models and models trained from scratch became smaller when more patients were used for training. Results of the transfer learning approach are visualized in Fig.~\ref{fig3}. Fig.~\ref{fig4} shows the error maps for both approaches on one example patient in large vessels (Fig.~\ref{fig4}A and C) and small vessels (Fig.~\ref{fig4}B and D). 

\section{Discussion}

We present a Wasserstein-GAN based model for the generation of synthetic TOF-MRA imaging data and corresponding labels. The model generated synthetic data of high quality, as evidenced visually and through the FID measure, and retained much of the predictive properties of the original images. Here, a predictive model for vessel segmentation trained on synthetic data alone showed a good performance on one dataset and excellent performance on an external validation set. The synthetic data were also successfully applied in a transfer learning approach where training was pre-initialized with weights from a model trained on synthetic data. It outperformed the models trained on real data. Our results mark a significant step towards the use of GAN-based models to generate synthetic and effectively anonymous data. Consequently, this approach has the potential to significantly accelerate research in the field of neuroimaging.

While the image-label pairs synthesized by the DCGAN showed some artifacts, the more recent GAN architectures (WGAN-GP and WGAN-GP-SN) produced higher resolution data that looked similar to the real data (Fig. 2). The superiority of the WGAN-approaches was confirmed by lower FID values as well as the improved performance of the U-net segmentation models trained on synthetic data. This can be explained by the inherent differences between Wasserstein-GANs and the DCGAN. In contrast to the DCGAN, the loss function of the WGAN-GP architectures utilizes the Earth Mover's distance and is bounded by a Lipschitz constraint (\cite{arjovsky_wasserstein_2017, gulrajani_improved_nodate}). This works as a robust regularization and enhances training stability while diminishing mode collapse at the same time. This explains why the WGAN-GP produced more realistic looking image-label pairs. Other studies confirm the superiority of Wasserstein GAN architectures over the DCGAN (\cite{arjovsky_wasserstein_2017, gulrajani_improved_nodate}). A recent addition to GAN architectures was the introduction of spectral normalization. This method additionally restricts the discriminator's weights for each layer in order to stabilize training even for high learning rates (\cite{miyato_spectral_2018}). As evidenced in our work, spectral normalization is also beneficial for the application of Wasserstein GANs, and the combination of both regularization techniques (WGAN-GP-SN) yielded the best image quality both by visual inspection as well as in terms of FID. These techniques have thus supported the preservation of the predictive properties for vessel segmentation within the synthetic patches. Therefore, it is likely that more sophisticated (future) GAN architectures will further improve the generation of synthetic data. Here, potential current candidate methods are progressive growing GAN (PG-GAN) or stacked GAN architectures (\cite{karras_progressive_2018, huang_stacked_2017}).

Whereas the data generated by WGAN-GP-SN consistently yielded the highest DSC in the transfer learning approach, this is not as apparent in other parts of the results. First, the 95HD did not show a consistent trend. Since the Hausdorff distance is vulnerable to outliers, we argue that it might not be as reliable as the DSC. This is also corroborated by the high standard deviation over the patients. Secondly, when training the U-net with real data and additional synthesized data (data augmentation), the performance only slightly increased for the WGAN-GP-SN. In addition, the DCGAN seems to perform slightly better. Here, the real training data used for the U-net was the same as for GAN training. Hence, the generated data contains information from the same underlying distribution and did not add much value. Further, the best performance by adding DCGAN generated data needs to be taken with caution and does not necessarily mean that this was the overall best performing generative model.

GAN architectures have the potential to generate anonymized data since the generator does not have direct access to the training data. This also holds true for this study: the generator synthesizes patch-label pairs from a noise vector. However, a recent study by \cite{hayes_logan_2019} shows that DCGANs might be vulnerable to so-called membership inference attacks  (\cite{shokri_membership_2017}). Such attacks aim to identify whether a given data sample was part of the original training set or not. To prevent this, differentially private GANs (DPGANs) have been introduced (\cite{xie_differentially_2018}). Here, carefully adjusted noise is introduced in the gradients during the discriminator's training. While these GANs have the potential to ensure a certain level of privacy, they show poorer performance to date (\cite{mukherjee_privgan_2020}) and have only been trained on natural image datasets yet. Training a DPGAN on sparse medical imaging datasets remains a major challenge. While DPGANs might provide even further advantages in anonymization, we argue that our synthesized patch-label pairs are effectively anonymized. For one, in the WGAN-GP-SN approach, we apply Lipschitz regularization techniques such as gradient penalty and spectral normalization.  \cite{wu_generalization_2019} found that these techniques might reduce information leakage and might even make the trained models resistant to membership inference attacks. Furthermore, we use randomly sampled 2D patches in this study. Thus, for a successful membership inference attack two events must coincide: First, the real training data that is protected by state-of-the-art hospital security systems has to be leaked. Second, the patches need to be extracted in the exact same way as in the GAN-training process to allow re-identification. The minuscule probability of these events to happen is comparable to other theoretical scenarios of state-of-the-art anonymization. For example, any tabular data anonymized using state-of-the-art techniques could be re-identified when compared with the leaked original data. Thus, we consider our generated patches anonymous and hence make them available for researchers upon request. 

Our results are also promising for AI in healthcare product development (\cite{Higgins_2020}). In the medical AI research setting, a strong focus on performance in homogeneous samples can be observed. This is in stark contrast to the requirements for a medical imaging product. A product is supposed to be used in a real world setting confronted with highly heterogeneous data reflecting different settings and multiple hardware options. Thus, product development should focus as much on training on heterogeneous data as on keeping the necessary performance (\cite{Higgins_2020}). This, however, is currently highly challenging as data is a scarce resource due to limited availability. Our results show that a relatively small amount of data is sufficient to generate robust results. Thus, a GAN-based anonymization approach could allow the generation of high quality data from a smaller number of patients from multiple locations that - in total - reflect the full distribution of soft- and hardware settings in the clinical setting. Here, the possibility to generate high-quality labels as evidenced by our study is also a great advantage. Notably, a GAN model also learns the quality of the labels provided during training. Thus, the final performance of any model trained on synthetic data will also be dependent on the quality of the real labels. Providing high-quality labels is no simple task and requires usually hours of manual labor by highly qualified medical staff. Thus, a novel GAN-based approach to product development could entail the high-quality labeling of relatively small data-sets from multiple data providers that are then anonymized and pooled for training. This would on one hand keep development costs relatively low which is a prerequisite for startup success. On the other hand, such an approach would ensure both high performance and low bias as the chance for out-of-sample data in the clinical setting would be significantly lowered.

Our study has several limitations. The DCGAN is 2D due to computational restrictions. 3D approaches could help extracting information about the 3D vessel tree structure and in this way improve the performance of the segmentation task. The computational restrictions also did not allow to try out more advanced GAN architectures such as PG-GAN. Another limitation is the calculation of the FID. Due to computational restrictions it was only calculated to confirm the quality of visually inspected images and not for every epoch in an end-to-end solution. Secondly, the FID for assessing the image quality might not be ideal. Although it is used as a quality measurement in the medical field (\cite{haarburger_multiparametric_2019, cao_auto-gan_2020}), it was originally designed for natural images and hence might not entirely capture relevant features for medical imaging. Thus, further research on assessing image quality specific to medical images should be undertaken.

\section{Conclusion}

This study marks a crucial step towards true anonymization of medical imaging data while maintaining essential predictive features within the image patch. We show that these features might be generalizable to another, independent dataset. Our initial performance for vessel segmentation on the PEGASUS dataset already is relatively high. We show that training more advanced GAN architectures can further increase the quality of synthesized image-label pairs. By using only one patient from a different cohort, we can achieve a high comparable performance on an independent dataset. Our synthesized image-label pairs allow other researchers to build models that only require few labeled patient data and will significantly facilitate research in this domain. It may be the case that our framework achieves similar results on other medical segmentation tasks. This could lead to a lower demand of labeled patient data and allow more data sharing of anonymized data. Nevertheless, further studies should assess the generalizability of this analysis to other (more complex) segmentation problems.

\section*{Disclosures}

Tabea Kossen reported receiving personal fees from ai4medicine outside the submitted work. Dr Madai reported receiving personal fees from ai4medicine outside the submitted work. Adam Hilbert reported receiving personal fees from ai4medicine outside the submitted work. 

While not related to this work, Dr Sobesky reports receipt of speakers honoraria from Pfizer, Boehringer Ingelheim, and Daiichi Sankyo. Furthermore, Dr Fiebach has received consulting and advisory board fees from BioClinica, Cerevast, Artemida, Brainomix, Biogen, BMS, EISAI, and Guerbet. Dr Frey reported receiving grants from the European Commission, reported receiving personal fees from and holding an equity interest in ai4medicine outside the submitted work. 

\section*{Acknowledgements}

This work has received funding by the German Federal Ministry of Education and Research through (1) the grant Centre for Stroke Research Berlin and (2) a Go-Bio grant for the research group PREDICTioN2020 (lead: DF).



%\begin{thebibliography}{99}
%\bibitem[\protect\citeauthoryear{Author}{2013}]{author2013}
%\bibitem[\protect\citeauthoryear{Jones}{2015}]{jones2015}
%Jones C.~D., 2015, Journal of Interesting Stuff, 17, 198
%\bibitem[\protect\citeauthoryear{Smith}{2014}]{smith2014}
%Smith A.~B., 2014, The Example Journal, 12, 345 (Paper I)
%\end{thebibliography}

%%%%%%%%%%%%%%%%%%%%%%%%%%%%%%%%%%%%%%%%%%%%%%%%%%

% Don't change these lines
%\bsp	% typesetting comment
\label{lastpage}
\end{document}